\documentclass[aps,prl,showpacs,superscriptaddress,twocolumn]{revtex4}
\usepackage{amsfonts}
\usepackage{txfonts}
\usepackage{amssymb}
\usepackage{amsbsy} 
\usepackage{epsfig}

\def\be{\begin{equation}} \def\ee{\end{equation}}
\def\bea{\begin{eqnarray}} \def\eea{\end{eqnarray}}

\def\nn{\nonumber}

\def\bsigma{{\boldsymbol \sigma}}
\def\bGamma{{\boldsymbol \Gamma}}
\def\bq{{\bf q}}
\def\bQ{{\bf Q}}
\def\bk{{\bf k}}
\def\br{{\bf r}}
\def\bl{{\bf l}}
\def\bp{{\bf p}}
\def\bE{{\bf E}}
\def\bB{{\bf B}}

\begin{document}
\title{Chiral anomaly, Charge Density Waves, and Axion Strings from Weyl Semimetals}

\author{Zhong Wang}

\affiliation{
Institute for Advanced Study, Tsinghua University, Beijing,  China, 100084}

\author{Shou-Cheng Zhang$^{1,}$}

\affiliation{
Department of Physics, Stanford University, Stanford, CA 94305}

\begin{abstract}

We study dynamical instability and chiral symmetry breaking in three dimensional Weyl semimetals, which turns Weyl semimetals into ``axion insulators''. Charge density waves (CDW) is found to be the natural consequence of the chiral symmetry breaking. The phase mode of this charge density wave state is identified as the axion, which couples to electromagnetic field in the topological $\theta{\bf E}\cdot{\bf B}$ term. One of our main results is that the ``axion strings'' can be realized as the (screw or edge) dislocations in the charge density wave, which provides a simple physical picture for the elusive axion strings. These axion strings carry gapless chiral modes, therefore they have important implications for dissipationless transport properties of Weyl semimetals with broken symmetry.

\end{abstract}

\maketitle

\emph{Introduction.} Topological insulators are among the most active
research fields in condensed matter physics
recently\cite{qi2010a,hasan2010,qi2011}. Among the remarkable aspects
of topological insulators is the ubiquitous role played by Dirac
fermions. In fact, most of the recently discovered topological
insulators can be regarded as massive Dirac fermion systems with
lattice regularization\cite{qi2011}. When the mass vanishes, we have
massless Dirac fermions, which are two copies of Weyl fermion with
opposite chiralities. The Dirac fermions, which obey the Dirac
equation, are described by spinors with four components, while the
Weyl fermions are two-component fermions described by the following
simple Weyl equation \bea \pm v_F\bsigma\cdot\bk \psi= E\psi
\label{weyl} \eea where $\pm$ is referred as chirality ($+$ for left
handed, $-$ for right handed), and $v_F$ is the Fermi velocity. Since
the Dirac fermion can be decomposed into two copies of Weyl fermion,
the latter is more elementary building block. In fact, it is our
current understanding of nature that the elementary fermions such as
quarks and electrons fall into the Weyl fermion framework because the
left-handed and right-handed fermions carry different gauge charges
in the standard model of particle physics\cite{weinberg1996}.

It is worth noting that in writing down Eq.(\ref{weyl}) we have
assumed that the two ``Weyl points'', at which the energy gap closes,
are both located at $\bk=0$. In the particle physics context, this
assumption seems to be natural, however, in condensed matter physics,
without imposing symmetry constraints such as time reversal symmetry
and inversion symmetry, Weyl points are generally located at
different points in the momentum space. This fact has interesting
consequences for Weyl fermion systems with broken symmetry, as we
will show in this paper.

Although Weyl fermion plays a crucial role in the description of
elementary fermions in nature,  it has been studied in the condensed
matter context only very
recently\cite{wan2011,volovik2003,burkov2011,zyuzin2012a,witczak2012,hosur2012,
aji2011,liu2012,
xu2011,yang2011,wang2012e,halasz2012,kolomeisky2012,jiang2012,delplace2012,meng2012,
garate2012,grushin2012,son2012}. Weyl semimetals in three dimensions
(3d) are analogous to graphene\cite{neto2009} in 2d in the sense that
both are described in terms of gapless fermions with approximately
linear dispersion, but the 3d Weyl semimetals are richer in that they
are more closely related to various fundamental phenomena such as the
chiral anomaly\cite{liu2012,son2012,aji2011}. Unlike the topological
insulators, whose transport is dominated by topologically protected
surface states, in the Weyl semimetal the bulk transport is most
important. Their unique semi-metallic behaviors in 3d can potentially
be engineered for semiconductor industry.

Interaction effect plays a fundamental role in the dynamics of Weyl
fermions. One possibility is the pairing interaction which leads to
the superconducting instability. Qi, Hughes and Zhang's fermi surface
topological invariant\cite{qi2010d} implies that if the pairing
amplitudes have opposite sign for Weyl points with the opposite
chirality, topological superconductors are obtained. Another
consequence of interaction, which we will focus in this paper, is the
spontaneous chiral symmetry breaking. Chiral symmetry breaking is the
phenomenon of spontaneous generation of an effective mass of Weyl
fermions, namely a paring between the fermions (electrons) and
anti-fermions (holes) with different chiralities. Due to chiral
anomaly, the Goldstone mode $\theta$ is coupled to electromagnetic
field as $\theta\bE\cdot\bB$, therefore, this Goldstone mode is an
``axion''\cite{peccei1977,wilczek1978,weinberg1978,wilczek1987}.

In this paper, we focus on the chiral symmetry breaking in Weyl
semimetal and axion strings in condensed matter context, which have
new features absent in particles physics.  We would like to mention
that axion string can be realized on the surface of topological
insulators with a magnetic domain wall\cite{qi2008}, and axionic
dynamics has been studied in topological magnetic
insulators\cite{li2010,wang2011c}. In the present paper we study a
new route to axionic dynamics through chiral symmetry breaking
induced by interaction effect. The resultant states are charge
density wave (CDW) states, which are experimentally observable. One
of our main results is that the (screw or edge) dislocations of CDW
are exactly the ``axion strings'', which are important topological
defects carrying gapless chiral modes. In these chiral modes
electrons move solely in one direction without backscattering. In
particles physics,  axion strings have interesting cosmological
implications such as the gravitational lenses
effect\cite{witten1985}, but observable evidences are elusive. Axion
strings in condensed matter systems have the advantage that they are
much easier to detect. In the Weyl semimetal studied in the present
paper, axion strings have important effects on the transport
properties since they provide chiral modes supporting dissipatonless
transport.

\emph{Dynamical chiral symmetry breaking.} We consider the simplest
model for dynamical chiral symmetry breaking, which nevertheless
captures the most salient physical consequences. First let us present
the free part $H_0$ of the Hamiltonian. The four-band model studied
here is a simplified version of the model of Weyl semimetal given in
Ref.\cite{burkov2011,liu2012}. We have $H_0=c^\dag h c$ with \bea h
=v_F\sum_{i=1}^{3}\Gamma^i (-i\partial_i -eA_i - q_i\Gamma) -eA_0
\eea where we have defined the Dirac matrices
$\Gamma^\mu$($\mu=0,1,2,3$) satisfying
$\{\Gamma^\mu,\Gamma^\nu\}=2\delta^{\mu\nu}$, $\Gamma$ is the
chirality operator with the properties $\Gamma^2=1$ and
$[\Gamma,\Gamma^i]=0$\cite{note_dirac}, $A_\mu$ is the external
electromagnetic potential, and $\bq=(q_1,q_2,q_3)$ is a vector that
shifts the gapless points away from $\bk=0$. The simplest choices of
the Dirac matrices are $\Gamma=\tau^3\otimes 1$ and
$\Gamma^i=\tau^3\otimes \sigma^i$ ( $i=1,2,3$ ). The low energy modes
from the left-hand ($\Gamma=+1$) chirality is described by $h_+(k)
\approx v_F  \bGamma\cdot(\bk-\bq) = v_F \bsigma\cdot (\bk-\bq)$ near
$\bk=\bq$. Similarly, there is a Weyl point at $-\bq$ for the
right-handed ($\Gamma=-1$) chirality. The low energy dynamics are
dominated by these two Weyl points located at $\bQ_1=\bq$ and
$\bQ_2=-\bq$ respectively, with  \bea  h_\pm(k) = \pm v_F
\bsigma\cdot (\bk-\bQ_i) \eea  where the prefactor $\pm 1$ is the
chirality. For later convenience let us define two operators
$\tau^\pm =\tau^1\pm i\tau^2$ with the property $\{\tau^\pm,
\Gamma\}=0$.

Now we would like to investigate the effects of four-fermion
interaction in Weyl semimetals. Let us write down the effective
action in the imaginary time as \bea S=\int d\tau d\br
\{c^\dag_\br[\partial_\tau + h + m^*(\br)\tau^+ + m(\br)\tau^- ]c_\br
+\frac{|m(\br)|^2 }{g}\} \label{h-s} \eea in which we have written
the interaction in terms of the auxiliary field $m(\br)$, which can
be integrated out to give the four-fermion interaction
$-g(c_\br^\dag\tau^+ c_\br)(c_\br^\dag\tau^- c_\br)$.

It usually happens that a dynamically generated energy gap can lower
the ground state energy of a nominally gapless system. Let us
investigate such possibility of condensation $\langle
m(\br)\rangle\neq 0$. Since the low energy dynamics are dominated by
the two Weyl points, let us write down the expansion
$c_{\br}=e^{i\bQ_1 \br}c_{L,\br} + e^{i\bQ_2 \br} c_{R,\br}+\cdots$,
where $c_{R/L}$ are cut off in the momentum space at $\Lambda$,  i.e.
$c_{L/R,\br} = \sum_{|\bp|<\Lambda} e^{i\bp\br}c_{L/R,\bp}+\cdots$.
and the ``$\cdots$'' terms are high energy modes with
$|\bp|>\Lambda$. At the mean field level we have $m(\br) = -g\langle
c^\dag_\br\tau^+ c_\br \rangle = -g e^{-i\bQ\br}\langle
c_L^\dag\tau^+ c_R\rangle $, where we have defined
$\bQ=\bQ_1-\bQ_2=2\bq$. We note that $\langle c_L^\dag\tau^+
c_R\rangle $ is a ``slow'' field whose characteristic momentum is
small compared to $|\bQ|$.

In the momentum space the fermion matrix in Eq.(\ref{h-s}) can be
approximated by $M=-i\omega+v_F\tau^3\bsigma\cdot\bp + m^*(\bQ)\tau^+
+ m(\bQ)\tau^-$ at low energy, therefore, we can obtain the gap
equation

\bea \frac{1}{2g} = \int \frac{d\omega d^3p}
{(2\pi)^4}\frac{1}{\omega^2 +v_F^2 p^2+|m|^2} \label{gap}\eea from
the mean-field relation $m(\br) = -g\langle c^\dag_\br\tau^+ c_\br
\rangle$. The solution to Eq.(\ref{gap}) can be obtained as
$\frac{1}{g_c}-\frac{1}{g} = \frac{1}{8\pi^2v_F^3}|m|^2
\ln\frac{v_F^2\Lambda^2+|m|^2}{|m|^2} $, where $g_c =
\frac{8\pi^2v_F}{\Lambda^2} $. We have taken a Lorentz-invariant
cutoff $|\bp|<\Lambda,\omega<v_F\Lambda$ in the above calculation,
but if we take the cutoff only for $|\bp|$ but not for the $\omega$,
we can check that $g_c$ takes the same value. Because we are
concerned with the cases with $|m|<<\Lambda$, the solution can be
approximated by \bea \frac{1}{g_c}-\frac{1}{g} =
\frac{1}{8\pi^2v_F^3}|m|^2 \ln\frac{v_F^2\Lambda^2}{|m|^2}
\label{gap2} \eea which shows that dynamical symmetry breaking (or
``exciton condensation'') occurs only when the interaction is
sufficiently strong ($g>g_c$).

A qualitative understanding of $g_c$ is simple. The kinetic energy
per fermion is $E_K\sim v_F\Lambda$, while the interaction energy per
fermion is $E_I \sim g\Lambda^3$, where $\Lambda^3$ accounts for the
spatial density of Weyl fermions. To have chiral condensation, we
must have $E_K\sim E_I$, or $g_c\sim v_F/\Lambda^2$. To satisfy this
condition, larger $E_I$ (stronger interaction) and smaller $E_K$
(narrower bandwidth in Dirac dispersion) is favored.

\emph{Axion dynamics and topological theta term.} We have seen in the
previous section that when $g>g_c$, chiral symmetry is spontaneously
broken. From symmetry consideration, the Ginzburg-Landau effective
action (omitting chiral anomaly at this stage) of $m(\br)$ can be
expressed as \bea S_{m} =\int dtd\br [\frac{1}{2}\gamma(|\partial_t
m'|^2 -v_a^2 |\partial_i  m'|^2)+ \delta |m'|^2 + \eta|m'|^4]
\label{gl}\eea where $\gamma,v_a,\delta,\eta$ are phenomenological
parameters, and $m' \equiv me^{i\bQ\cdot\br}$ is the ``slow'' field.
In the symmetry breaking phase, $\delta<0$ and $|m|$ develops a
nonzero expectation value. The effective action $S_m$ is invariant if
we shift the phase of $m(\br)$ by a spacetime-independent phase
factor, but in fact this symmetry is broken by chiral anomaly, which
endows a topological theta term to the effective action, as we
explain below. Let us first write $m(\br)=|m(\br)|\exp[-i\bQ\cdot\br
- i\theta(\br)]$. We can perform a chiral transformation
$c(\br)\rightarrow c(\br)e^{-i(\bQ\cdot\br+\theta)\Gamma/2}$, then
$m(\br)\rightarrow m(\br) e^{i(\bQ\cdot\br+\theta)}$. After this
chiral transformation the phase of $m(\br)$ is removed and $m(\br)$
becomes real numbers, however,  due to the fact that the fermion path
integral measure is not invariant\cite{fujikawa1979}, this chiral
transformation generates an anomalous term $S_{{\rm anomaly}}
=\frac{e^2}{32\pi^2}\int dtd\br
\epsilon^{\mu\nu\lambda\rho}(\bQ\cdot\br+\theta)
F_{\mu\nu}F_{\lambda\rho} = \frac{e^2}{4\pi^2}\int dtd\br
(\bQ\cdot\br+\theta){\bf E}\cdot{\bf B}$, where we have used the
natural unit  $\hbar=c=1$.

Taking the above chiral anomaly into account, the fluctuations of
$\theta$ is described by the following simple axionic effective
action \bea S_{\theta}=&&\frac{f_a^2}{2}\int
dtd\br[(\partial_t\theta)^2-v_a^2(\partial_i\theta)^2] \nn \\ && +
\frac{e^2}{4\pi^2}\int dtd\br (\bQ\cdot\br+\theta){\bf E}\cdot{\bf B}
\label{theta} \eea where the notation ``$f_a$''($\equiv\gamma|m|$) is
deliberately chosen because it is analogous to the pion decay
constant $f_\pi$, namely that $f_a$ is the ``axion decay constant''.
We can also define a normalized field $a=f_a\theta$ and put
Eq.(\ref{theta}) into a more standard form \bea  S_a = &&
\frac{1}{2}\int dtd\br [(\partial_t a)^2-v_a^2(\partial_i a)^2] \nn
\\ && + \frac{e^2}{4\pi^2}\int dtd\br (\bQ\cdot\br +
\frac{a}{f_a}){\bf E}\cdot{\bf B} \label{theta_a} \eea There is an
effective action analogous to the last term for pion-photon coupling
in high energy context, which is responsible for the famous
two-photon decay of neutral pion. The axion-photon coupling is
proportional to $1/f_a\sim 1/(\gamma |m|)$, thus we have the
counterintuitive conclusion that when the chiral condensation goes
weaker, the axion-photon coupling becomes stronger.

Various topological responses can be calculated from the effective
action given in Eq.(\ref{theta}). Taking derivative with respect to
$A_\mu$, we have the current \bea j^\mu =
\frac{e^2}{8\pi^2}\epsilon^{\mu\nu\lambda\rho}(Q_\nu+\partial_\nu\theta)
F_{\lambda\rho} =
\frac{e^2}{8\pi^2}\epsilon^{\mu\nu\lambda\rho}(Q_\nu+\frac{\partial_\nu
a}{f_a}) F_{\lambda\rho} \label{response} \eea Analogous topological
responses have been studied in topological insulators\cite{qi2008},
in which the first term in absent. Let us consider the special case
of a constant magnetic field $B_z{\bf z}$ along the ${\bf z}$
direction, then the charge density given by Eq.(\ref{response}) is
\bea j^0=\frac{e^2}{4\pi^2}(Q_z+\partial_z\theta) B_z =
\frac{e^2}{4\pi^2}(Q_z+\frac{\partial_z a}{f_a})B_z
\label{density}\eea The first term here is readily understood as
layered quantum Hall effects\cite{bernevig2007a}, with layer
thickness $2\pi/|\bQ|$. The second term can be understood as follows.
Let us consider the case $\bQ=0$ for simplicity, and take $|m|=0$
first. In a constant magnetic field $B_z{\bf z}$, the dispersions of
Weyl fermions can be obtained as $E_n(p_z)=\pm v_F\sqrt{ p_z^2+2eB_z
n}$ with $n=0,1,\cdots$. The two gapless modes are the $n=0$ Landau
levels with $E(p_z)=\pm v_F p_z$, where $\pm$ corresponds to left and
right chirality respectively. Now a mass term $m=|m|e^{i\theta}$
mixes the two counter-propagating (essentially 1d) modes and opens a
gap. The 1d charge density  can be obtained from the
Goldstone-Wilczek formula\cite{goldstone1981} $j^0
|_{1d}=\frac{1}{2\pi}\partial_z\theta$, therefore, the final result
of 3d charge density is $j^0=(eB_z/2\pi)(\partial_z\theta/2\pi)=
\frac{e^2}{4\pi^2}B_z\partial_z\theta$, where we have added the
density of states $eB_z/2\pi$ of Landau levels. This is exactly the
second term of Eq.(\ref{density}).

\emph{Phase of charge density wave is the dynamical axion.} Now we
will show that the chiral symmetry breaking leads to density waves,
among which the CDW is the simplest. The charge density is given by
\bea \rho_1(\br) = \langle c_{\br}^\dag \tau^1 c_{\br}\rangle
=-\frac{ m(\br)+m^*(\br)}{2g} =-\frac{|m|}{g}\cos(\bQ\cdot\br+\theta)
\label{rho-1}\eea and similarly $\rho_2(\br) = \langle c_{\br}^\dag
\tau^2 c_{\br}\rangle =i[m(\br)-m^*(\br)]/2g=
|m|\sin(\bQ\cdot\br+\theta)/g$. Let us explain their physical
consequences. In fact, they depend on the physical degree of freedom
to which $\tau$ is referred. Let us take a simplest example, namely
that $\tau^3=\pm 1$ refers to $(|A\rangle\pm|B\rangle)/\sqrt{2}$,
where $A,B$ refer to two inequivalent sites in a unit
cell\cite{note_parity}, then $\tau^1=\pm 1$ refers to $A/B$ site,
thus $\rho_1=\rho_A-\rho_B$ is the staggered CDW. If we look at the
charge density on site $A$ (or $B$), it shows an oscillation with
wavelength $2\pi/|\bQ|$.  In more general cases, other density
modulation, such as CDW of more general types and spin density waves
can show up.

The natural question is how to experimentally detect the CDW. Apart
from bulk measurements, it can also be detected by simpler surface
measurements such as scanning tunneling microscope (STM). Denoting
the angle between the surface normal and $\bQ$ as $\alpha$, we can
obtain the surface CDW wavelength as
$\lambda_{2d}=\frac{2\pi}{|\bQ||\sin\alpha|}$.

It is worth noting that the interaction effect and CDW was studied in
2d Dirac systems in Ref.\cite{raghu2008,weeks2010,wang2010c}. More
recently, interaction effect on the surface of weak topological
insulators has been studied in Ref.\cite{liu2011}, in which CDW has
also important physical consequences. The relation to chiral anomaly
is absent in these studies because the systems considered there are
in 2d, where the concept of chirality is lacking.

\emph{Dislocations in charge density wave are axion strings.} Let us
turn to the central part of this paper, namely the identification of
CDW dislocations as axion strings, which may provide dissipatonless
chiral transport channel in 3d bulk materials. An axion string $l$ is
a one dimensional dislocation of axion field, around which the axion
field $\theta$ changes by $2\pi$, namely that \bea \int_C d\theta
=2\pi \eea where $C$ is a small contour enclosing $l$ clockwise.
Axion strings are closely related to chiral anomaly, as was studied
long ago in the work by Callan and Harvey\cite{callan1985} in the
particle physics context. In the Weyl semimetals studied in the
present paper, the axion strings have clear  geometrical picture
because  $\theta$ is exactly the phase of CDW. More explicitly,
suppose that $\bQ=(Q_x,Q_y,Q_z)=(0,0,Q)$, then it follows  from
Eq.(\ref{rho-1}) that the peaks of CDW are located at 2d planes
$(x,y,z_n)$ with \bea z_n= -\frac{\theta+2\pi n}{Q}\label{z}\eea
where $n=$integer. When $\theta$ is shifted, the peak position $z_n$
follows the shifting of $\theta$. In fact, the shifting of $z_n$
around the small loop $C$ enclosing the axion string is readily
obtained from Eq.(\ref{z}) as \bea \int_C dz_n =-\frac{\int_C
d\theta}{Q} =-\frac{2\pi}{Q}\eea which is exactly the wavelength of
the CDW. The Burgers vector of the axion string as a dislocation of
CDW is exactly $(0,0,-2\pi/Q)$. For a general CDW wave vector $\bQ$,
the Burgers vector is readily obtained as $-2\pi\bQ/|\bQ|^2$.

Let us refer to the orientation of the axion string $l$ as $\hat{\bl}$.
According to the relative orientation of  $\hat{\bl}$ and $\bQ$, the
axion string appears as different types of dislocation. When
$\hat{\bl}$ is parallel with $\bQ$, we have a screw dislocation
[Fig.(a)], while when $\bQ$ is perpendicular with $\hat{\bl}$, we
have an edge dislocation [Fig.(b)]. We would like to mention that the edge dislocation with chiral modes has been studied in
Ref.\cite{teo2010}. Weyl semimetals provide a natural route to
realize such interesting topological defects. In the cases of edge
dislocation, the origin of chiral modes is most clear, because we can think of them as the edge states of a 2d quantum Hall system, which is just the slice appearing as the ``edge''.

There are chiral modes propagating along the axion
strings\cite{callan1985}, therefore, axion strings may serve as
unique transport channels in a 3d materials with axionic dynamics.
Such chiral modes carry dissipationless current just like the quantum
Hall edge and quantum anomalous Hall edge states, but the former are
distinct in that they are buried in 3d bulk.

It is worth mentioning that dislocations in topological materials has
also been studied in Ref.\cite{ran2009}, but we would like to
emphasize several prominent differences between the axion strings
studied here and the dislocation lines in weak topological insulators
studied in Ref.\cite{ran2009}. First, in Ref.\cite{ran2009} the
dislocations carrying gapless modes are indeed dislocations of
crystal lattice, while in our paper the crystal lattice remains
intact, and axion strings are just dislocations of CDW. Second, the
gapless modes studied in Ref.\cite{ran2009} are helical modes, which
are unstable towards back scattering if time reversal symmetry is
broken, while the gapless modes living on the axion strings studied
in the present paper are robust chiral modes. It is also worth noting
that in Ref.\cite{liu2012} line dislocation with chiral modes was
studied, but CDW is absent there, more importantly, the bulk is also
gapless there and the coupling between dislocation mode and bulk mode
can induce dissipation.

To conclude this section we remark that the formation of axion
strings in Weyl semimetals can be triggered by rapidly lowering the
temperature from $T>T_c$ to $T<T_c$, where $T_c$ is the critical
temperature of chiral condensation (Kibble-Zurek mechanism).

\begin{figure}
\includegraphics[width=9.0cm, height=4.0cm]{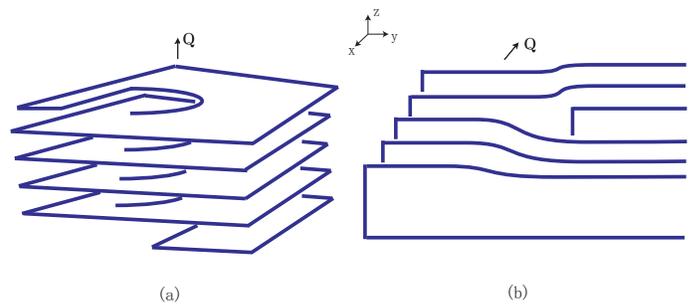}
\caption{Axion strings as dislocations of charge density wave.
(a)Screw dislocation. (b)Edge dislocation. The delineated sheets are
the peaks of CDW. In both (a) and (b), the axion string is along the
${\bf z}$ direction. The Burgers vector is parallel to the axion
string in (a), while it is perpendicular to it in (b).} \label{string}
\end{figure}

\emph{Discussions and Conclusions.} We have studied the dynamical
chiral symmetry breaking and topological responses in Weyl
semimetals. We have adopted a simple four-fermion interaction to
simplify formulas. In more realistic models $g$ is replaced by
$g(\bQ)$. We note that an attractive interaction $-g(\bQ)<0$ at
momentum $\bQ$ is needed for the chiral symmetry breaking. The values
of $g(\bQ)$ for various materials depend on the material details, which
is beyond the scope of this paper.  It is useful to mention that an
effectively attractive electron-electron interaction can appear at
some special momenta commensurate with the reciprocal lattice.  Such
electron-lattice coupling effect is responsible for the Peierls
transitions in 1d systems, and we expect that dynamical chiral
symmetry breaking may also occur in 3d by this mechanism if
$2\pi/|\bQ|$ is commensurate with the crystal lattice. In this case
the chiral symmetry breaking can be thought of as generalized Peierls
transitions, which induces dimerization, trimerization, etc.

Our model provides a geometrical picture of axion, which manifests
itself as the phase of CDW.  One of our main results is the
identification of axion strings as CDW (edge or screw) dislocations,
which has no analog in particle physics. The axion strings have 1d
robust chiral modes along them, which have great potential
applications if the chiral symmetry breaking (exciton condensation)
of Weyl fermion is realized in experiment. In this paper we studied
the general cases with $\bQ\neq 0$. When $\bQ=0$ there is no CDW
associated with the chiral symmetry breaking, but the axion strings
do exist and have important implication for 3d transport properties.


ZW thanks Chao-Xing Liu and Xiao-Liang Qi for helpful discussions. ZW
is supported by Tsinghua University Initiative Scientific Research
Program (No. 20121087986). SCZ is supported by the NSF under grant
numbers DMR-0904264 and the Keck Foundation.

\emph{Note added.} Recently we became aware of a
related work\cite{zyuzin2012} on symmetry breaking in Weyl semimetal
by Zyuzin and Burkov, though CDW and axion strings were not studied.
Due to nonzero density of states considered in their work, the gap
equation is also different from ours.

\appendix

\bibliography{axion}

\begin{thebibliography}{46}
\bibitem[{not({\natexlab{a}})}]{note_dirac}
\bibinfo{howpublished}{Note that we have used the nonrelativistic notation in
  which the Dirac matrices $\Gamma^\mu$ can be written in the relativistic
  notation $\gamma^\mu$ as $\Gamma^i = \gamma^0\gamma^i$($i=1,2,3$) and $\Gamma
  = \gamma^5$, therefore $[\Gamma^i,\Gamma]=0$ is consistent with
  $\{\gamma^i,\gamma^5\}=0$.}
\bibitem[{not({\natexlab{b}})}]{note_parity}
\bibinfo{howpublished}{If there is an inversion symmetry exchanging $A$ and $B$
  site, then $\tau^3$ is just the parity eigenvalue.}
\end{thebibliography}
\end{document}